\documentclass{article}
\usepackage{iclr2027_conference,times}

\usepackage{amsmath,amssymb}
\usepackage{booktabs}
\usepackage{graphicx}
\usepackage{wrapfig}
\usepackage{capt-of}
\usepackage{microtype}
\usepackage{multirow}
\usepackage{xcolor}
\usepackage{xurl}
\usepackage{algorithm}
\usepackage[noend]{algpseudocode}
\usepackage{pifont}
\usepackage{tikz}
\usepackage{pgfplotstable}
\pgfplotsset{compat=1.18}
\usetikzlibrary{arrows.meta,positioning,fit,calc}
\usepackage{hyperref}
\hypersetup{hidelinks}
\usepackage{enumitem}
\usepackage{caption}

\usepackage{fontawesome5}

\DeclareMathOperator*{\argmax}{arg\,max}
\DeclareMathOperator*{\argmin}{arg\,min}

\newcommand{\specialcell}[2][c]{%
	\begin{tabular}[#1]{@{}c@{}}#2\end{tabular}}

\newcommand{\myvspace}[1]{\vspace{#1}}

\algnewcommand\algorithmicinput{\textbf{Input:}}
\algnewcommand\Input{\item[\algorithmicinput]}
\algnewcommand\algorithmicoutput{\textbf{Output:}}
\algnewcommand\Output{\item[\algorithmicoutput]}

\newcommand{\method}{\textsc{Cobalt}}

\title{\method{}: Leveraging Expert Co-activation for Efficient Distributed MoE Training}

\author{\textbf{Junkang Zhou}$^{1}$~~~\textbf{Xinyi Liu}$^{2}$~~~\textbf{Fangcheng Fu}$^{3\,\text{\faEnvelope}}$\\
$^{1}$Zhejiang University, $^{2}$Peking University, $^{3}$Shanghai Jiao Tong University\\
$^{\text{\faEnvelope}}$\texttt{ccchengff@sjtu.edu.cn}}

\begin{document}
\maketitle
\suppressfloats[t] 

\begin{abstract}
Mixture-of-Experts (MoE) has increasingly become a mainstream approach for scaling large language models, as it expands model capacity while keeping computation cost nearly constant. 
Training large-scale MoE models relies on Expert Parallelism (EP), which distributes expert replicas across GPUs and exchanges tokens through all-to-all communication. 
The efficiency of EP is often constrained by two system bottlenecks: cross-node token transfers are limited by inter-node bandwidth, while skewed expert workloads lead to imbalanced computation across GPUs.
 Prior work mitigates these bottlenecks based on per-expert workload statistics, but overlooks the fact that experts could share the communication. 

In this work, we empirically present the observation that many pairs of experts are frequently co-activated by individual tokens. Motivated by this, we present \method{}, an efficient MoE training framework that leverages expert co-activation to reduce cross-node traffic and workload imbalance. 
\method{} adopts a two-stage expert layout planner that adapts expert layout to the evolving expert co-activation and workload conditions.
It periodically co-locates frequently co-activated experts on the same node to reduce the cross-node communication, and performs per-step intra-node adjustment to rebalance the workloads. 
Subsequently, we develop a communication-aware task assignment method that routes tokens to fewer remote nodes based on the current expert layout.
Experiments on 32 B200 GPUs show that \method{} achieves up to 1.53-2.41$\times$ (1.28-1.89$\times$ on average) of speedup compared to existing MoE training frameworks, while reducing cross-node token traffic by 75.74\%-99.26\%.
\end{abstract}

\section{Introduction}
\label{sec:intro}

Mixture-of-Experts (MoE) has become the de facto architecture for scaling large language models (LLMs). By routing each token to a small subset of experts, it increases the model scale while keeping per-token computation nearly constant~\citep{shazeer2017outrageously,switch}.
Currently, frontier open-source LLMs have been scaled to hundreds of billions or trillions of parameters through the MoE architecture~\citep{deepseekv4,glm5,kimik3}.

\begin{figure}[!t]
\begin{minipage}{0.50\linewidth}
\vspace{0pt}
\centering
\begingroup
\definecolor{flowA}{HTML}{F4B968}
\definecolor{flowB}{HTML}{91C6E8}
\definecolor{flowC}{HTML}{A5D0A4}
\definecolor{flowD}{HTML}{C6B4DF}
\definecolor{flowRoute}{HTML}{B4DCD5}
\definecolor{flowComm}{HTML}{F4A0A8}
\begin{tikzpicture}[x=1cm,y=1cm,
  font=\rmfamily\fontsize{7.5}{8.5}\selectfont,
  every node/.style={inner sep=1.3pt},
  arrow/.style={-{Stealth[length=1.3mm]},draw=black!70,line width=.45pt},
  stage/.style={draw=black!55,fill=black!4,minimum width=1.32cm,
    minimum height=.31cm,inner sep=1pt},
  expert/.style={draw=black!65,minimum width=.49cm,minimum height=.40cm},
  replica/.style={draw=black!65,minimum width=.49cm,minimum height=.40cm}]
  \foreach \x/\nodeIndex in {0/0,3.35/1} {
    \draw[black!55,densely dotted,rounded corners=3pt]
      (\x,0.22) rectangle ++(3.13,4.35);
    \node[font=\rmfamily\bfseries\fontsize{8}{9}\selectfont]
      at (\x+1.565,4.40) {Node \nodeIndex};
  }
  \foreach \x/\gpu/\owner/\copy/\ownerColor/\copyColor in {
    .82/0/a/b/flowA/flowB,2.32/1/c/a/flowC/flowA,
    4.17/2/b/c/flowB/flowC,5.67/3/d/a/flowD/flowA} {
    \node at (\x,4.05) {GPU \gpu};
    \node[stage] (att\gpu) at (\x,3.65) {Attention};
    \node[stage] (gate\gpu) at (\x,3.13) {Gating};
    \node[stage,fill=flowRoute] (copy\gpu) at (\x,2.61) {Assignment};
    \draw[arrow] (att\gpu) -- (gate\gpu);
    \draw[arrow] (gate\gpu) -- (copy\gpu);
    \draw[arrow] (copy\gpu) -- (\x,2.22);
    \node[expert,fill=\ownerColor] at (\x-.28,1.58) {$\owner$};
    \node[replica,fill=\copyColor] at (\x+.28,1.58) {$\copy$};
    \draw[arrow] (\x,1.91) -- (\x,1.80);
    \draw[arrow] (\x,1.35) -- (\x,1.23);
    \node[stage,align=center,minimum height=.45cm] (sum\gpu) at (\x,.60) {Weighted\\[-1pt]sum};
    \draw[arrow] (\x,.94) -- (sum\gpu);
  }
  \node[stage,fill=flowComm,minimum width=6.25cm]
    at (3.24,2.08) {Dispatch (all-to-all communication)};
  \node[stage,fill=flowComm,minimum width=6.25cm]
    at (3.24,1.08) {Combine (all-to-all communication)};
  \node[expert,fill=flowA] at (.40,-.15) {$a$};
  \node[anchor=west] at (.72,-.15) {Replica for expert $a$};
  \node[stage,fill=flowRoute,minimum width=.49cm] at (4.14,-.15) {};
  \node[anchor=west] at (4.47,-.15) {Task assignment};
\end{tikzpicture}
\endgroup
\captionof{figure}{\small{Illustration of expert parallelism and replication. In the example, experts $a, b, c, d$ have 3, 2, 2, and 1 replica(s), respectively.}}
\label{fig:moe-computation-flow}
\end{minipage}
\begin{minipage}{0.01\linewidth}
$ $
\end{minipage}	
\begin{minipage}{0.48\linewidth}
  \centering
  \begingroup
  \usepgflibrary{fpu}
  \definecolor{breakAttention}{HTML}{A5B7DA}
  \definecolor{breakComputation}{HTML}{D0D0D0}
  \definecolor{breakCommunication}{HTML}{F0D6A0}
  \pgfplotstableread[col sep=comma,row sep=\\]{
attention_ms,moe_computation_ms,communication_ms,timing_scale\\
865.771524,416.970451,2674.283617,2\\
7529.062197,3632.705153,15538.332266,0.5\\
2335.118924,1395.267608,6465.505530,1\\
885.701232,502.639771,5800.833002,2\\
7496.403890,3775.173361,42055.189260,0.5\\
2215.081951,983.156473,15606.901589,1\\
  }\baselineData
  \begin{tikzpicture}[x=1cm,y=.73cm,
    font=\rmfamily\bfseries\fontsize{7.5}{8.5}\selectfont,
    every node/.style={inner sep=1.3pt}]
    \foreach \x/\y/\tone/\name in {
       .12/1.30/breakAttention/Others,
      2.30/1.30/breakComputation/{MoE computation},
      .12/.78/breakCommunication/{MoE communication}} {
      \filldraw[fill=\tone,draw=black!55,line width=.4pt]
        (\x,\y-.08) rectangle ++(.23,.16);
      \node[anchor=west] at (\x+.29,\y) {\name};
    }
    \draw[black!65,line width=.4pt] (.75,.10) -- (.75,-4.00) -- (6.15,-4.00);
    \foreach \y/\model in {.12/Hunyuan3,-1.33/GLM-4.5-Air,-2.78/DeepSeek-V3}
      \node[anchor=west,fill=white,font=\rmfamily\bfseries\itshape\fontsize{8}{9}\selectfont]
        at (.88,\y) {\model};
    \foreach \row/\y/\ep in {0/-.36/16,3/-.82/32,1/-1.81/16,4/-2.27/32,2/-3.26/16,5/-3.72/32} {
      \node[anchor=east,text=black!70] at (.65,\y) {EP\ep};
      \pgfkeys{/pgf/fpu=true,/pgf/fpu/output format=fixed}
      \pgfplotstablegetelem{\row}{timing_scale}\of\baselineData
      \pgfmathsetmacro{\timingScale}{\pgfplotsretval}
      \pgfplotstablegetelem{\row}{attention_ms}\of\baselineData
      \pgfmathsetmacro{\attentionMs}{\pgfplotsretval}
      \pgfplotstablegetelem{\row}{moe_computation_ms}\of\baselineData
      \pgfmathsetmacro{\computationMs}{\pgfplotsretval}
      \pgfplotstablegetelem{\row}{communication_ms}\of\baselineData
      \pgfmathsetmacro{\communicationMs}{\pgfplotsretval}
      \pgfmathsetmacro{\totalMs}{\attentionMs+\computationMs+\communicationMs}
      \pgfmathsetmacro{\communicationPercent}{100*\communicationMs/\totalMs}
      \pgfmathsetmacro{\attentionEnd}{\timingScale*\attentionMs/1000}
      \pgfmathsetmacro{\computationEnd}{\timingScale*(\attentionMs+\computationMs)/1000}
      \pgfmathsetmacro{\communicationEnd}{\timingScale*\totalMs/1000}
      \pgfkeys{/pgf/fpu=false}
      \foreach \start/\finish/\tone in {
        0/\attentionEnd/breakAttention,
        \attentionEnd/\computationEnd/breakComputation,
        \computationEnd/\communicationEnd/breakCommunication} {
        \filldraw[fill=\tone,draw=black!55,line width=.4pt]
          ({.75+.18*\start},\y-.12) rectangle
          ({.75+.18*\finish},\y+.12);
      }
      \node[anchor=west] at ({.75+.18*\communicationEnd+.06},\y)
        {\pgfmathprintnumber[fixed,precision=0]{\communicationPercent}\%};
    }
    \node at (3.45,-4.80) {Training time};
  \end{tikzpicture}
  \endgroup
  \myvspace{-10pt}
  \caption{\small{Time breakdown of training three MoE models with EP degrees of 16 and 32. The value next to each bar indicates the portion of MoE communication in training time.}}
  \label{fig:step-breakdown}
\end{minipage}
\myvspace{-12pt}
\end{figure}

\begin{figure}[!t]
\centering
\includegraphics[width=\linewidth]{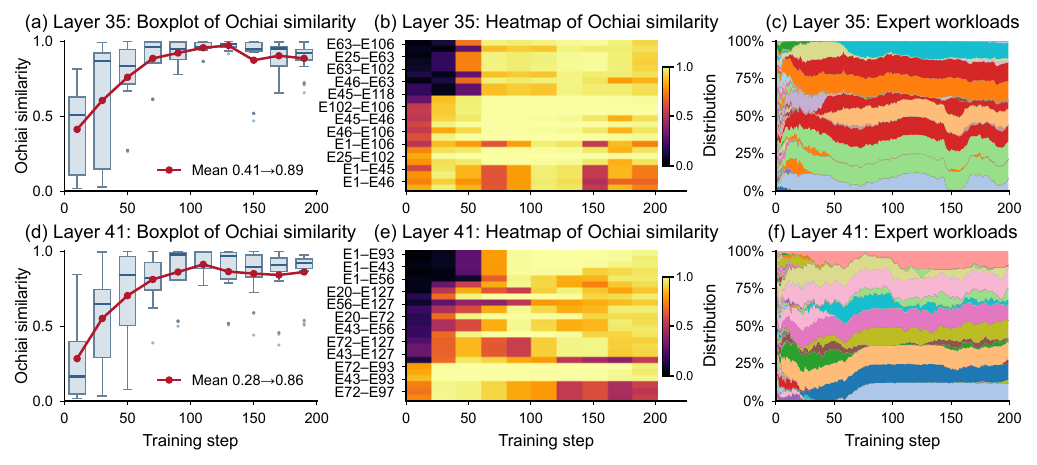}
\myvspace{-15pt}
\caption{\small{Observations when training GLM-4.5-Air. Statistics are collected on two model layers. (a)\&(d) Ochiai similarity generally increases as training progresses. (b)\&(e) The expert co-activation condition of individual expert pairs changes during training. (c)\&(f) Distribution of expert workloads w.r.t. training steps.}}
\label{fig:observation}
\myvspace{-10pt}
\end{figure}

In a nutshell, for each token in an input sequence, the MoE architecture uses a gating module to select $K$ experts for computation, and merges the results from these experts. 
Expert parallelism (EP)~\citep{gshard,switch} is the standard way for the distributed training of large MoE models.
As depicted in Figure~\ref{fig:moe-computation-flow}, EP distributes experts across GPU ranks to amortize the memory and computation. 
Since the experts may be selected by different numbers of tokens, a common practice is to replicate experts with high workloads, while the placement of all experts' replicas jointly affects the workload balance across GPU ranks. 
Given each token's top-$K$ expert selection, the task assignment process chooses one replica for each token-expert computation task.
Then, EP uses all-to-all communication to dispatch tokens to their assigned replicas for computation and triggers another all-to-all communication to combine the results.

However, all-to-all communication of EP can dominate training time as it spans more GPUs.
For example with Figure~\ref{fig:step-breakdown}, our profiling results show that MoE communication accounts for 58\%-83\% of the training time across three models, and this portion increases as the EP degree grows.
Similar results have also been reported in prior works~\citep{smartmoe,liu2026laermoe}.
Such a high communication cost can be largely attributed to the costly cross-node token transfers due to the bandwidth hierarchy. 
For instance, an NVIDIA B200 GPU server provides up to 1.8 TB/s of NVLink bandwidth for GPUs within the same node, whereas the inter-node InfiniBand connection provides at most 400 GB/s of bandwidth in aggregate~\citep{nvidia_dgxb200}. 
This gap has driven communication libraries like DeepEP~\citep{deepep2025} to deduplicate token transfers at the node level. 
Specifically, when two selected experts reside on different GPUs of the same remote node, the token is only sent across the inter-node connection once, and is forwarded via the intra-node link. 
Consequently, cross-node traffic depends on the number of distinct remote nodes selected by a token. 

Although there exist various efforts to improve MoE communication and workload balance~\citep{he2022fastermoe,smartmoe,liu2026laermoe,flexmoe}, they are mainly driven by per-expert workload statistics.
However, with node-level token transfer deduplication, an expert's workload alone cannot indicate whether it could share the cross-node communication with another expert. 
Particularly, the potential savings of cross-node traffic depend on whether two experts are simultaneously activated by many tokens. 
To investigate whether such potential savings can be realized, we analyze the expert selections from the perspective of expert pairs, and observe the \textit{expert co-activation} phenomenon.
Specifically, during the training of GLM-4.5-Air~\citep{glm45}, we track the set of active tokens for each expert in each step, and compute the Ochiai similarity\footnote{Given two sets $A,B$, their Ochiai similarity is $|A\cap B|/\sqrt{|A||B|}$, which is widely used to measure the similarity between sets. In our observation, a high Ochiai similarity of an expert pair indicates that the two experts are co-activated by many tokens.} of every expert pair that has co-activation. 
Figure~\ref{fig:observation}(a)(d) visualizes two layers' growth in expert co-activation: the average Ochiai similarity rises from 0.41 to 0.89 and from 0.28 to 0.86, respectively. 
It indicates that expert co-activation exists in many expert pairs, and occurs more frequently as training progresses.

Inspired by this, we aim to leverage expert co-activation to accelerate expert parallelism training. 
However, doing so faces two essential challenges. 
First, as shown in Figure~\ref{fig:observation}(c)(f), different experts vary substantially in terms of their workloads (i.e., the number of token-expert computation tasks). 
Simply placing frequently co-activated experts on one node may concentrate token-expert tasks there, causing stragglers. 
Thus, it calls for reducing cross-node traffic while maintaining workload balance. 
Second, the co-activated expert pairs change during training, as depicted in Figure~\ref{fig:observation}(b)(e), and so does each expert's workload (Figure~\ref{fig:observation}(c)(f)). 
As a result, we need to adaptively adjust the expert layout (including expert replication and replica placement) according to the evolving conditions, while ensuring the overhead does not harm overall training efficiency, which is unexplored yet.

To address these challenges, we present \method{}, an efficient distributed MoE training framework that utilizes the expert co-activation information to reduce cross-node communication while taking workload balance into account.
\begin{itemize}[leftmargin=*]
\item 
To begin with, we formulate the joint optimization problem, which aims to deduce the expert layout and token-expert task assignment to minimize the communication cost of cross-node token transfers under the workload balance constraint.
\item 
Subsequently, we dissect the problem to determine expert layout and task assignment accordingly. 
For expert layout, we propose a two-stage planner, which periodically updates the layout globally based on historical expert workloads and co-activation information, and performs per-step intra-node rebalancing to adjust the expert replicas within each node. 
Such a two-stage design adapts the expert layout to the evolving co-activation statistics while limiting the overhead.
For task assignment, we develop a communication-aware task assignment method, which favors fewer remote nodes for each token given its top-$K$ expert selections and the current expert layout.
\item 
We implement \method{} and conduct experiments on 32 NVIDIA B200 GPUs. Empirical results show that \method{} outperforms existing works by up to 1.53-2.41$\times$ (1.28-1.89$\times$ on average), and shrinks the cross-node token transfers by 75.74\%-99.26\%. 
\end{itemize}

In summary, we make three major contributions. 
\textit{(New perspective)} To our knowledge, this is the first work to leverage expert co-activation for cross-node communication reduction and workload balance in distributed MoE training. 
\textit{(New framework)} We develop \method{}, which features a two-stage expert layout planner and a communication-aware task assignment method to reduce cross-node communication while maintaining workload balance.
\textit{(State-of-the-art performance)} Empirical results on 32 NVIDIA B200 GPUs show that \method{} achieves up to 1.53-2.41$\times$ (1.28-1.89$\times$ on average) speedup compared to existing training frameworks.

\section{Related Work}
\label{sec:related}

\textbf{Load-aware placement and replication.}
Existing systems adapt either data placement or expert placement to routing demand. 
FasterMoE~\citep{he2022fastermoe}, SmartMoE~\citep{smartmoe}, FlexMoE~\citep{flexmoe} and LAER-MoE~\citep{liu2026laermoe} adjust expert replication and replica placement in response to load imbalance.
SYMI~\citep{symi2026} and Themis~\citep{themis} address where expert state resides and when it is materialized.
These methods are driven by per-expert load statistics, and overlook finer-grained routing statistics.
Another line of research exploits such finer-grained statistics.
NetMoE~\citep{liu2025netmoe} profiles the expert load of each training sample and rearranges sample placement between layers to optimize cross-node communication.
HierMoE~\citep{hiermoe} implements node-level token transfer deduplication, and swaps expert positions to minimize the cross-node communication time.
Occult~\citep{occult2025} clusters experts with similar workloads onto one GPU rank via one-shot offline profiling, and then keeps the placement fixed.
These methods reduce communication cost, but do not address the computation imbalance across GPUs.
In contrast, \method{} exploits expert co-activation while accounting for workload balance across GPUs.

\textbf{MoE gating and specialization.}
MoE architectures and gating modules shape which experts a token activates: 
DeepSeekMoE~\citep{deepseekmoe}, Mixtral~\citep{mixtral}, and OLMoE~\citep{olmoe} develop expert specialization and gating designs; 
BASE~\citep{base} and Expert Choice~\citep{expertchoice} balance expert selections.
TA-MoE~\citep{tamoe2022} and LocMoE~\citep{locmoe2024} introduce topology or locality into gating.
Occult~\citep{occult2025} changes the selected experts through modified fine-tuning. 
\method{} adapts expert layout and task assignment during training without changing the gating module's selected experts or weights.

\textbf{Communication and kernel optimization.}
HierMoE~\citep{hiermoe} reduces redundant token transfers through hierarchical deduplication and uses expert swaps to balance communication load.
DeepEP~\citep{deepep2025} likewise supports node-level token deduplication. 
Lina~\citep{lina2023}, Lancet~\citep{lancet2024}, Comet~\citep{comet2025}, and FLUX~\citep{flux} schedule or overlap communication with computation, while FastMoE~\citep{he2021fastmoe}, Tutel~\citep{tutel}, MegaBlocks~\citep{megablocks}, and PIT~\citep{pit} improve MoE execution and sparse computation.
\method{} builds on node-level token deduplication, and is orthogonal to the scheduling and kernel optimizations above.

\textbf{Inference scheduling.}
Semantic Parallelism~\citep{li2026semantic} groups experts according to their activation patterns and schedules requests for MoE inference. 
Libra~\citep{yang2026libra} predicts expert activations to hint replication and token assignment.
These systems exploit activation patterns for inference placement and scheduling. 
\method{} targets training, where model updates can change co-activation patterns and expert replication additionally requires correct gradient aggregation, optimizer updates, and parameter consistency.

\section{Method}
\label{sec:mone-method}

As discussed in Section~\ref{sec:intro}, there exists the expert co-activation phenomenon during the training of MoE models. 
\method{} leverages this to reduce cross-node communication
while balancing computation across GPUs. 
Table~\ref{tab:notation} summarizes the frequently used notations.

\begin{wrapfigure}{r}{0.5\textwidth}
\begin{minipage}{0.5\textwidth}
\myvspace{-10pt}
  \centering
  \caption{\small{Frequently used notations.
   }}
  \label{tab:notation}
  \small
  \begin{tabular}{ll}
    \toprule
    $E$ & Number of experts in an MoE layer. \\
    $K$ & Number of experts selected per token. \\
    $G$ & Number of ranks for EP (i.e., EP degree). \\
    $N$ & Number of nodes for EP. \\
    $\texttt{N}(r)$ & Node containing rank $r$. \\
    $T$ & The number of tokens per training step. \\
    $s_t$ & Source rank of token $t$. \\
    $\mathcal{D}_t$ & The set of selected experts of token $t$. \\
    $\mathcal{P}_r$ & The set of experts with a replica on rank $r$. \\
    $r_{te}$ & The rank assigned token-expert task $(t,e)$. \\
    $w_{er}$ & Workload of expert $e$'s replica on rank $r$. \\
    $\rho$ & Budget to control the maximum number of \\&replicas per rank. \\
    $\varepsilon$ & Tolerance factor of workload imbalance. \\
    \bottomrule
  \end{tabular}
\end{minipage}
\end{wrapfigure}

\subsection{Problem Formulation and Design Overview}
\label{sec:mone-formulation}

\textbf{System decisions.}
We consider one MoE layer with $E$ experts distributed over
$G$ GPU ranks on $N$ nodes, with $G/N$ ranks per node. 
For each token $t$, the set of $K$ selected experts is denoted as $\mathcal{D}_t$.
\method{} preserves these selections, and determines
where the resulting token--expert computation tasks execute.
Specifically, there are two decisions to make.
\textit{(1) Expert layout} comprises 
the \textit{expert replication} that determines how many replicas each expert has, and 
the \textit{replica placement} that determines the GPU rank holding each replica. 
\textit{(2) Task assignment}
determines the GPU rank that each token--expert computation task is assigned to.
In a nutshell, the expert layout defines the available ranks for each task,
and task assignment selects one of those ranks. 
Thus, they jointly affect communication and load balance: sending a
token's tasks to fewer nodes saves network transmission but concentrates workload on these nodes, whereas distributing
them facilitates workload balance but causes higher cross-node communication.

\textbf{Problem formulation.}
Formally, the expert layout is represented by $\mathcal{P}=\{\mathcal{P}_r\}$, where
$\mathcal{P}_r$ is the set of experts who have a replica on rank $r$.
These sets describe both expert replication and replica placement:
expert $e$ has $\sum_r\mathbb{I}[e\in \mathcal{P}_r]$ replicas in total.
To avoid out-of-memory errors, we enforce a replica budget $\rho\ge0$, which allows at most $(1+\rho)E/G$ replica slots
per rank.\footnote{$\rho$ controls the number of replica slots per rank.
Every token--expert computation task executes once, so a higher $\rho$ (more
replica slots) allows an expert's work to be divided across ranks while
consuming more memory.}
For task assignment, we denote $r_{te}$ as the rank that executes task $(t,e)$, where $e\in \mathcal{D}_t$ and $e\in \mathcal{P}_{r_{te}}$.
We formulate the problem as searching for the expert layout and task assignment that minimize communication while taking workload imbalance into account.

For communication, we focus on cross-node communication cost 
as inter-node bandwidth is smaller than
intra-node bandwidth.
It is common practice to deduplicate token transmission at the node level~\citep{deepep2025}: 
if two ranks on the same node need to receive the same token, then it is done by one cross-node transmission followed by an intra-node transmission (a.k.a. intra-node forwarding).
Thus, the communication cost can be modeled as 
the total number of cross-node token transfers, i.e.,
\begin{equation}
  \textstyle V=\sum_{t=1}^{T}
    \left|\bigcup_{e\in \mathcal{D}_t}\left\{\texttt{N}(r_{te})\right\} \setminus \{\texttt{N}(s_t)\}\right|,
  \label{eq:mone-traffic}
\end{equation}
where $T$ is the number of tokens, $\texttt{N}(r)$ denotes the node that rank $r$ resides, $\bigcup_{e\in \mathcal{D}_t}\left\{\texttt{N}(r_{te})\right\}$ is the set of destination nodes for token $t$, and $s_t$ is the source rank for token $t$.

For workload imbalance, we model the workload of the replica of expert $e$ on rank $r$ as the number of tasks assigned to it, i.e., $w_{er}=\sum_{t: e\in \mathcal{D}_t}\mathbb{I}[r_{te}=r]$.
Since every task executes once, we have $\sum_r \sum_e w_{er}=KT$, and the average per-rank workload is $KT/G$.
To restrict workload imbalance, 
we bound each rank's load by $(1+\varepsilon)$ times the average,
where $\varepsilon\ge0$ controls the allowed imbalance. 

Putting them together, we formulate the following problem:
\begin{equation}
\begin{aligned}
  \textstyle \argmin_{\{\mathcal{P}_r\},\,\{r_{te}\}}\quad & V \\
  \text{s.t.}\quad
    & \textstyle\sum_r\mathbb{I}[e\in \mathcal{P}_r]\ge1, \,\forall e,
    \quad && |\mathcal{P}_r|\le (1+\rho)E/G, \,\forall r,\\
    & e\in \mathcal{P}_{r_{te}}, \,\forall t,\ e\in \mathcal{D}_t,
    && \textstyle\sum_{e\in \mathcal{P}_r} w_{er}\le(1+\varepsilon)KT/G, \,\forall r.
\end{aligned}
\label{eq:mone-reference}
\end{equation}
The first two constraints ensure every expert has at least one replica 
while the number of replicas per rank does not exceed the budget. 
The third constraint ensures that each task is assigned to an available expert replica, and
the last constraint is for workload balance.

\begin{figure}[!t]
\centering
\includegraphics[width=\linewidth]{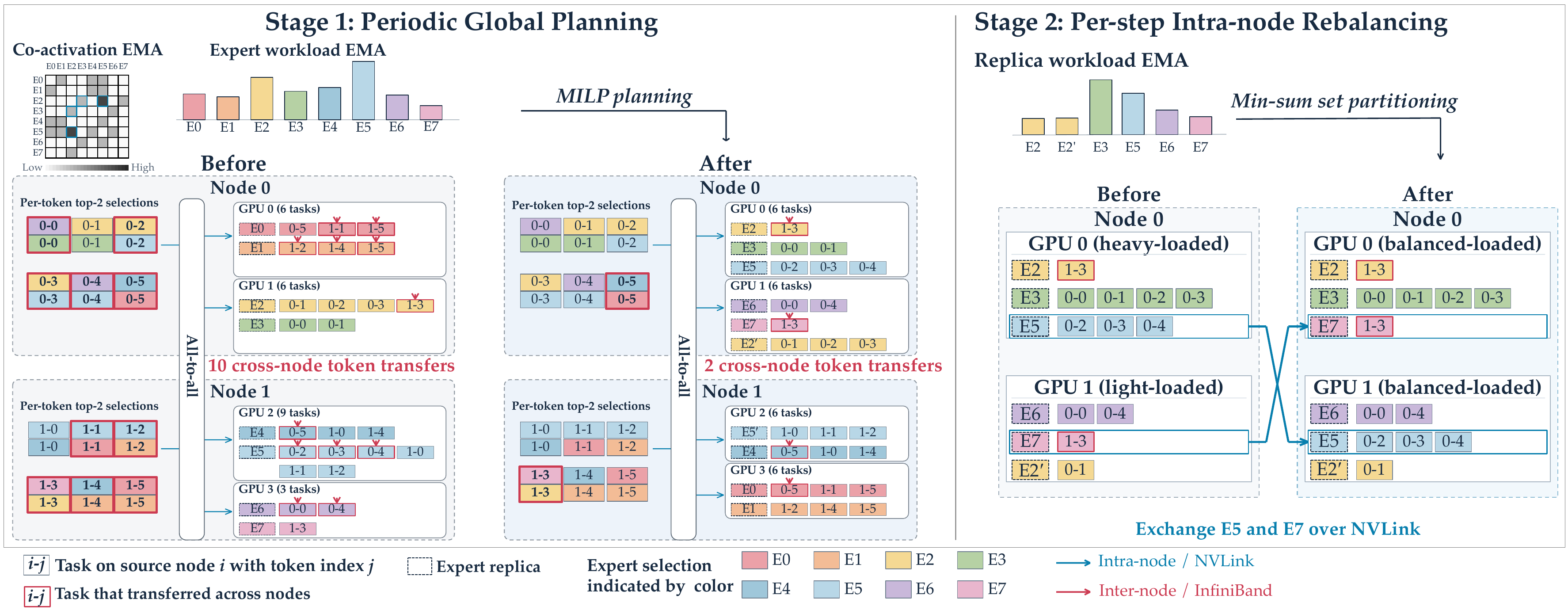}
\myvspace{-15pt}
\caption{\small{Overview of \method{}'s two-stage expert layout planning.}}
\label{fig:overview_planning}
\myvspace{-15pt}
\end{figure}

\textbf{Design overview.}
Solving Equation~\eqref{eq:mone-reference} gives us the optimal expert layout and task assignment given the top-$K$ selections of one training step.
However, each step's top-$K$ selections become available only
after gating. Computing and applying a new expert layout for every
step would require frequent cross-node transfers of expert parameters.
We therefore separate expert layout planning from task assignment.
First, the \emph{two-stage expert layout planner} (Section~\ref{sec:mone-layout})
uses historical statistics for periodic global updates and rebalances
GPUs within each node at every training step.
Then, the \emph{communication-aware task assignment}
(Section~\ref{sec:mone-routing}) then assigns the current tasks
over the current expert layout, balancing GPU loads while favoring
shared token transfers. Lastly, Section~\ref{sec:mone-implementation}
describes how these decisions are applied during training.

\subsection{Two-stage Expert Layout Planner}
\label{sec:mone-layout}

As mentioned in Section~\ref{sec:intro}, the co-activation evolves during training, 
which indicates that the optimal expert layout is not fixed. 
However, adjusting the expert layout globally requires moving
parameters across the inter-node links, incurring high communication cost. 
In contrast, exchanging replicas between ranks on the same node can utilize the faster intra-node links.
This motivates us to design a 
two-stage expert layout planner.
As depicted in Figure~\ref{fig:overview_planning},
it periodically updates both expert replication and replica placement 
to re-new the expert layout across all nodes, 
and adjusts replica placement within each node for each training step. 
By doing so, we can limit cross-node expert parameter transfer overhead 
while allowing per-step load rebalancing.

\textbf{Stage 1: Periodic global planning.}
To adapt to the changes in co-activation, we periodically re-plan the expert layout 
(including expert replication and replica placement)
while taking the expected workload balance into account. 
To capture the workload assigned to each expert as well as the co-activation phenomenon, 
the planner uses two statistics from previous training steps:
$\hat{w}_e$ is the exponential moving average (EMA)
of expert $e$'s assigned computation task count. 
$\hat{c}_{ij}$ is the EMA of the number of tokens selecting both experts $i$ and $j$, representing how frequently they are co-activated.

Since we do not have the per-token information 
(i.e., top-$K$ selections $\mathcal{D}_t$ and source rank $s_t$) for periodic planning, 
we transform Equation~\eqref{eq:mone-reference} for expert layout planning. 
For one thing, we introduce auxiliary decision variables $w^\prime_{er}$ to denote 
the expected workload assigned to the replica of expert $e$ on rank $r$. 
For another, we turn the objective from minimizing cross-node communication to 
maximizing the total co-activation. 
Based on these, we re-formulate the following problem:
\begin{equation}
\begin{aligned}
  \textstyle \argmax_{\{\mathcal{P}_r\},\,\{w^\prime_{er}\}}\quad
    & \textstyle\sum_n\sum_{i<j}\hat{c}_{ij}\,
      \mathbb{I}[i\in \mathcal{P}^{(n)} \wedge j\in \mathcal{P}^{(n)}]\\
  \text{s.t.}\quad
    & \textstyle\sum_r\mathbb{I}[e\in \mathcal{P}_r]\ge1, \,\forall e,
    && |\mathcal{P}_r|\le (1+\rho)E/G, \,\forall r\\
    & \textstyle\sum_r w^\prime_{er}=\hat{w}_e, \,\forall e
    && w^\prime_{er}=0\quad\text{if }e\notin \mathcal{P}_r, \\
    & \textstyle\sum_{e\in \mathcal{P}_r} w^\prime_{er}\le(1+\varepsilon)KT/G, \,\forall r
\end{aligned}
\label{eq:mone-planning}
\end{equation}
Here, $\mathcal{P}^{(n)} = \cup_{r: \texttt{N(r)}=n} \mathcal{P}_r$ 
denotes the set of experts that have replica(s) residing on node $n$.
The first two constraints are the same as Equation~\eqref{eq:mone-reference}. 
The third and fourth constraints ensure all computation tasks are properly assigned. 
And the last constraint prevents workload imbalance.

Equation~\eqref{eq:mone-planning} can be re-written as a mixed-integer linear programming (MILP) problem. 
(Due to the space constraint, we leave the detailed MILP formulation in Appendix~\ref{sec:appendix-milp}.) 
Therefore, it can be efficiently solved via existing libraries like SCIP~\citep{scip}.
Moreover, the solving can be overlapped with the training process. 
The solution of Equation~\eqref{eq:mone-planning} consists of 
the expert layout ($\{\mathcal{P}_r\}$) and the expected workload division across ranks ($\{w^\prime_{er}\}$). 
The former will be used to update the expert layout, 
while the latter will be utilized in the subsequent stage.

\textbf{Stage 2: Per-step intra-node rebalancing.}
Between global planning, \method{} supports adjusting replica placement 
within each node  to achieve load rebalancing. 
Similarly, we estimate each
replica's workload by tracking the EMA of the number of
token-expert computation tasks executed by that replica. 
Unlike global planning, the EMA here is at the replica level rather than expert level, and is initialized from the solution of periodic planning 
(i.e., $w^\prime_{er}$ of Equation~\eqref{eq:mone-planning}) 
and updated after each training step. 
Subsequently, for each node, we need to 
partition the replicas among the ranks, 
while minimizing the highest workload among the ranks. 

Undoubtedly, this is a variant of the min-sum set partitioning problem, 
with two additional constraints that two replicas of the same expert 
cannot reside on one rank, and that 
each rank contains at most $(1+\rho)E/G$ replicas. 
We solve this problem via a greedy bin-packing approach, 
which sorts the replicas by the EMA workloads and partitions them into sets greedily. 
Although it may not achieve the optimal replica placement, 
such a heuristic approach is extremely fast and can be overlapped with 
the model computation (detailed in Section~\ref{sec:mone-implementation}), 
which is necessary since it is done per-step.

\subsection{Communication-aware Task Assignment}
\label{sec:mone-routing}

Ideally speaking, we could keep the expert layout ($\{\mathcal{P}_r\}$) fixed and 
solve Equation~\eqref{eq:mone-reference} to deduce the current batch's task assignments ($\{r_{te}\}$). 
However, the top-$K$ selections become available only
after gating, placing this solving on the critical path. 
Moreover, even with a fixed layout, the problem retains $KT$ 
decision variables, which makes the solving prohibitively time-consuming.  
Therefore, we seek a heuristic task assignment approach that reduces communication based on the current expert layout, 
rather than directly solving Equation~\eqref{eq:mone-reference} for every training step.

Our key observation is that a token's cross-node communication cost
depends on its destination nodes, while choosing ranks within those
nodes does not change the number of cross-node input transfers.
We assign tasks in two steps: first selecting the fewest remote
nodes needed to serve each token, and then sampling a replica within
each assigned node. 
The routine is shown in Algorithm~\ref{alg:mone-router}.

\textbf{Step 1: Selecting execution nodes.}
Let $\mathcal R_{e,n}=\{r\in\mathcal R_n:e\in \mathcal{P}_r\}$ be the ranks
holding expert $e$ on node $n$, and let
$\mathcal N_e=\{n:\mathcal R_{e,n}\ne\varnothing\}$ be the nodes holding expert $e$. 
For token $t$, selected experts that are currently on the source node
will be executed there, requiring no cross-node input transfer.
The remaining experts form
$\mathcal U_t=\{e\in \mathcal{D}_t:\texttt{N}(s_t)\notin\mathcal N_e\}$ (line 3).
To minimize the cross-node communication, 
we only need to solve a minimum set-cover problem, 
i.e., finding the smallest set of remote nodes 
that covers the remaining experts.
We solve this problem by enumerating node subsets with bitmasks.
Specifically, for each $e\in\mathcal U_t$, 
an $(N-1)$-bit mask $m_{t,e}$ is able to mark
all remote nodes holding replicas of expert $e$ (line 4).
A candidate mask $b$ is feasible exactly when 
$b\mathbin{\&}m_{t,e}\ne0,\forall e\in\mathcal U_t$,
and the feasible mask with the smallest population count 
is the best solution to the minimum set-cover problem (lines 5-8). 
Finally, for each $e\in\mathcal U_t$, 
we randomly select an execution node $n_{te}$
from the nodes indicated by the best feasible mask (lines 9-10).

\begin{wrapfigure}{r}{0.52\textwidth}
\myvspace{-20pt}
\begin{minipage}{0.52\textwidth}
\centering
\begin{algorithm}[H]
\captionof{algorithm}{\small{Communication-aware task assignment.}}
\label{alg:mone-router}
\small
\begin{algorithmic}[1]
\Input Per-token selected experts $\mathcal{D}_t$ and source rank $s_t$, per-expert residing ranks $\mathcal R_{e,n}$ and nodes
  $\mathcal N_e$
\Output Execution rank $r_{te}$ for each task $(t,e)$ 
\For{each token $t$ \textbf{in parallel}}
\State // \textit{Step 1: Select execution nodes.}
\State $\mathcal U_t\gets\{e\in \mathcal{D}_t:\texttt{N}(s_t)\notin\mathcal N_e\}$
\State $m_{t,e}\gets\operatorname{RemoteMask}(\mathcal N_e,\texttt{N}(s_t))$,
  $\forall e\in\mathcal U_t$
\State $b^\star\gets 2^{N-1}-1$ // \textit{Start from all remote nodes}
\For{$b =0,1,\cdots,2^{N-1}-2$}
  \If{$b\mathbin{\&}m_{t,e}\ne0, \forall e\in\mathcal U_t$ \textbf{and} $\operatorname{BitCount}(b)<\operatorname{BitCount}(b^\star)$}
      \State $b^\star\gets b$ // \textit{Fewer remote nodes but feasible}
  \EndIf
\EndFor
\For{each $e\in \mathcal U_t$}
  \State $n_{te}\gets\operatorname{RandomNode}(b^\star\mathbin{\&}m_{t,e})$
\EndFor
\For{each $e\in \mathcal{D}_t \setminus \mathcal U_t$}
  \State $n_{te}\gets \texttt{N}(s_t)$ // \textit{Execute on the local node}
\EndFor
\State // \textit{Step 2: Select execution ranks.}
\For{each $e\in \mathcal{D}_t$}
  \State $r_{te}\gets\operatorname{UniformChoice}(\mathcal R_{e,n_{te}})$
\EndFor
\EndFor
\State \Return $\{r_{te}:e\in \mathcal{D}_t\}$
\end{algorithmic}
\end{algorithm}
\end{minipage}
\myvspace{-15pt}
\end{wrapfigure}

\textbf{Step 2: Selecting execution ranks.}
For each task $(t,e)$, 
denote $n_{te}$ as its execution node selected in Step 1. 
We uniformly sample an execution rank
$r_{te}$ from $\mathcal R_{e,n_{te}}$ (lines 14-15). 
Thus, replicas of the same expert
within a node receive equal workloads in expectation.
This exactly matches our per-step layout adjustment: 
intra-node rebalancing distributes the replica workloads across
ranks, while this sampling distributes incoming tasks among the
available replicas.

It is noteworthy that our task assignment 
does not explicitly enforce a workload balance constraint. 
However, this constraint is already considered during
periodic expert layout planning and per-step intra-node rebalancing
(Section~\ref{sec:mone-layout}). 
And the uniform sampling in task assignment maintains the balance.

The complexity of Algorithm~\ref{alg:mone-router} is $O(T2^{N-1})$. 
In practice, the EP degree is not high, 
so $2^{N-1}$ can be viewed as constant 
(e.g., EP degree of $G=32$ gives $2^{N-1}=8$ since each node contains 8 GPUs). 
Additionally, we can parallelize the task assignment 
for the $T$ tokens. 
Consequently, the time cost of our task assignment process 
is negligible. 

\subsection{Implementation}
\label{sec:mone-implementation}

We implement \method{} on top of PyTorch. 
To deduplicate token transmission and improve communication efficiency, 
we adopt DeepEP~\citep{deepep2025} as the backend for the all-to-all communication. 
Both the MILP problem (Eq.~\eqref{eq:mone-planning}) of periodic global planning and 
the min-sum set partitioning problem of per-step intra-node rebalancing 
are solved on the CPU, 
whereas we implement a GPU kernel for the task assignment (Algorithm~\ref{alg:mone-router}) using Triton. 
By default, we trigger the global expert layout update 
every 10 training steps. 
The solving of expert layout is overlapped with 
model computation, 
while the global update necessitates pausing the training process 
and transmitting the expert model parameters of across nodes. 
However, the overhead is worthwhile given its benefits.
For each training step, we perform the intra-node load balancing 
immediately after parameter prefetching so that 
the adjustment can be fully overlapped by model computation.

\section{Experiments}
\label{sec:eval}

\subsection{Experimental Setup}
\label{sec:eval-setup}

\textbf{Setup.}
To evaluate the effectiveness of \method{}, we conduct experiments on 4 NVIDIA B200 GPU servers (32 GPUs in total) with 1.8 TB/s intra-node NVLink and 400GB/s inter-node InfiniBand connections. 
We compare \method{} with three baselines.
The first is a standard PyTorch implementation that applies EP to the MoE-FFN module and Fully Sharded Data Parallel (FSDP)~\citep{fsdp} to the other modules. 
We refer to this baseline as EP+FSDP. 
The other two baselines are SmartMoE~\citep{smartmoe}, and LAER-MoE~\citep{liu2026laermoe}. 
To be fair, all four frameworks use DeepEP~\citep{deepep2025} for the all-to-all communication in EP. 
For the evaluation workloads, we consider model architectures from Hunyuan3~\citep{hunyuan3}, GLM-4.5-Air~\citep{glm45}, and DeepSeek-V3~\citep{deepseekv3}. 
Due to the memory constraint, we restrict the number of layers to be 16, 45, and 12 for the three models, 
comprising 60B, 106B, and 140B model parameters, respectively. 
Since our goal is to improve system efficiency, we conduct the training on the C4 dataset~\citep{raffel2020t5}, which is widely used in LLM training, and set the training sequence length to 8192. 
In our experiments, all frameworks start from the same checkpoint of each model, which is obtained after 200 steps of training from scratch. 
By default, we set $\rho=0.5$ and $\varepsilon=$0.15. 
More details of the hardware and workload configurations are provided in Appendix~\ref{sec:appendix-eval}.

\textbf{Metrics.}
We mainly focus on training step time and the corresponding speedup. 
The training step time is averaged over 50 steps and includes the time cost of expert layout adjustment. 
We also report two metrics to interpret the efficiency: 
the cross-node communication volume caused by token transfers (excluding those caused by FSDP and expert layout update), 
and the workload imbalance ratio (computed as the maximum workload of a single rank divided by the average workload of all ranks). 
We make five runs for each experiment and report the mean and standard deviation.

\subsection{Experimental Results}
\label{sec:eval-results}

\begin{figure}[!t]
\centering
\includegraphics[width=\linewidth]{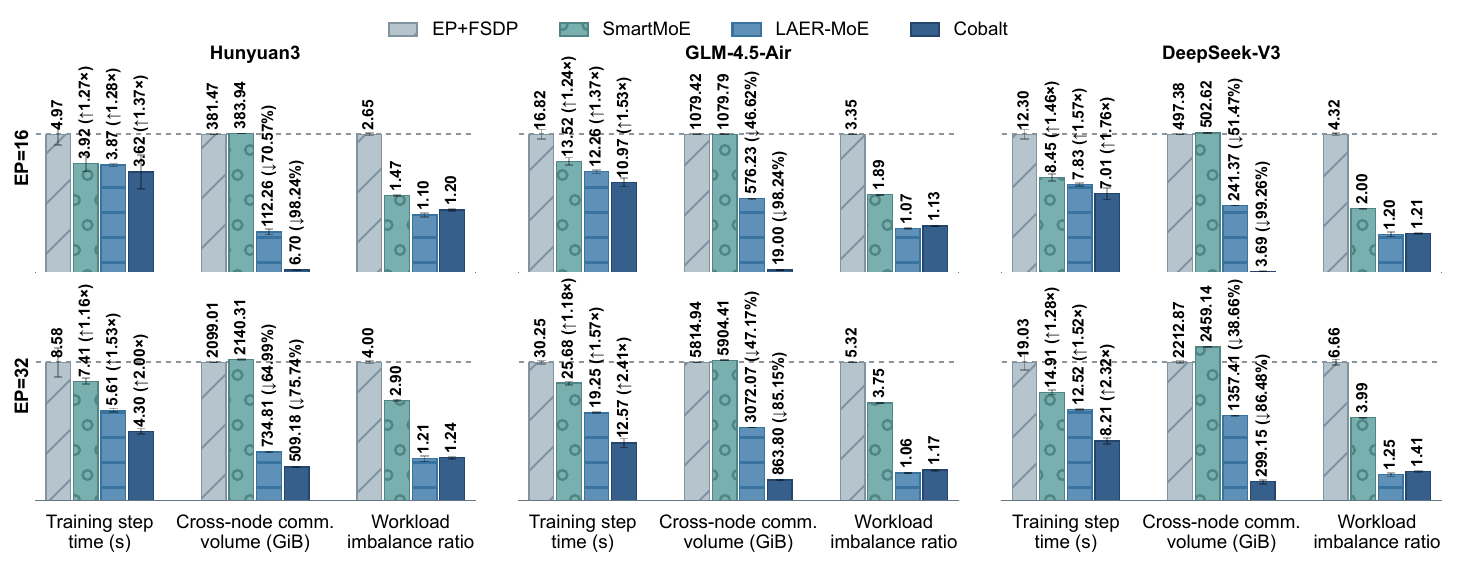}
\myvspace{-15pt}
\caption{\small{End-to-end comparison. We evaluate with 16 and 32 GPUs, respectively. For training step time, we provide the speedup compared to EP+FSDP on each bar. For cross-node communication volume, we provide the relative reduction compared to EP+FSDP.}}
\label{fig:eval_e2e}
\myvspace{-10pt}
\end{figure}

\textbf{End-to-end comparison.}
We first evaluate the end-to-end training efficiency. 
As shown in Figure~\ref{fig:eval_e2e}, \method{} consistently achieves the best efficiency, outperforming EP+FSDP, SmartMoE, and LAER-MoE by up to 2.41$\times$, 2.04$\times$, and 1.53$\times$ (1.89$\times$, 1.52$\times$, and 1.28$\times$ on average), respectively. 
Moreover, it can be observed that the speedup of EP=32 is higher than that of EP=16, which is reasonable because a larger EP size often leads to more cross-node communication.

To interpret the performance, we record the cross-node communication volume and workload imbalance ratio. 
The gain of \method{} stems from the reduction in both cross-node communication and workload imbalance. 
To be specific, \method{} significantly reduces cross-node communication volume by 75.74-99.26\% compared to EP+FSDP and consistently achieves the lowest level of workload imbalance. 
LAER-MoE mainly optimizes workload balance across ranks but does not address the cross-node communication bottleneck. 
In contrast, \method{} accounts for both, reducing cross-node communication volume by 30.71-78.29\% compared to LAER-MoE under a comparable level of workload ratio.
SmartMoE requires each expert either to have only one replica on a single rank or to be fully replicated across all ranks, so its performance falls behind ours, especially for EP=32.

\begin{figure}[!t]
\begin{minipage}[t]{0.49\textwidth}
\vspace{0pt} 
\centering
\includegraphics[width=\linewidth]{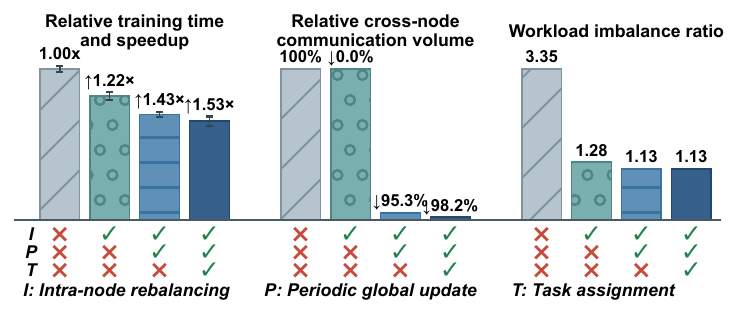}
\myvspace{-20pt}
\captionof{figure}{\small{Ablation studies (GLM-4.5-Air, EP=16).}}
\label{fig:eval_ablation}
\end{minipage}
\begin{minipage}[t]{0.01\textwidth}
$ $
\end{minipage}	
\begin{minipage}[t]{0.49\textwidth}
\vspace{0pt} 
\captionof{table}{\small{Overhead of layout planning and adjustment in one training step (GLM-4.5-Air).}}
\label{tab:eval_overhead}
\myvspace{-10pt}
\setlength{\tabcolsep}{3pt}
\resizebox{\linewidth}{!}{
\begin{tabular}{lcccc}
\toprule
& \multicolumn{2}{c}{Planning (seconds)} & \multicolumn{2}{c}{Adjustment (seconds)} \\
\cmidrule(lr){2-3} \cmidrule(lr){4-5}
& Global & Intra-node & Global & Intra-node \\
\midrule
EP16 & $1.21 {\scriptstyle \pm 0.43}$ & $0.25 {\scriptstyle \pm 0.12}$ & $2.04 {\scriptstyle \pm 0.06}$ & $0.03 {\scriptstyle \pm 0.01}$ \\
EP32 & $2.61 {\scriptstyle \pm 1.25}$ & $0.37 {\scriptstyle \pm 0.18}$ & $2.41 {\scriptstyle \pm 0.14}$ & $0.05 {\scriptstyle \pm 0.02}$ \\
\bottomrule
\end{tabular}
}
\end{minipage}
\myvspace{-15pt}
\end{figure}

\textbf{Ablation studies.}
We assess the effectiveness of the proposed techniques in \method{}. 
The results are provided in Figure~\ref{fig:eval_ablation}.
To begin with, our intra-node rebalancing contributes to workload balancing. It lowers the imbalance ratio from 3.35 to 1.28, yielding 1.22$\times$ of speedup in training time. 
Then, our periodic global update significantly shrinks the cross-node communication volume by 95.3\%, 
and it also helps decrease imbalance ratio to 1.13, 
resulting in a speedup of 1.43$\times$. 
Lastly, enabling our communication-aware task assignment 
can further reduce the cross-node communication by 98.2\%. 
Putting them together, \method achieves a speedup of 1.53$\times$ compared to the EP+FSDP baseline.

\textbf{Overhead of proposed techniques.}
We measure the overhead of our two-stage expert layout planner in Table~\ref{tab:eval_overhead}. 
In short, the overhead is minor for both planning and adjustment. 
Intra-node rebalancing takes negligible time and can be overlapped with previous layer's model computation. 
The global layout planning takes relatively longer time and increases w.r.t. the EP degree, yet it is still minor compared to the training time.
Although the global layout adjustment cannot be overlapped with model computation, it is triggered periodically, so the impact on overall training efficiency is limited. 
In addition, since our communication-aware task assignment is fully parallelized in a GPU kernel, its overhead stays within 1 millisecond, which is omittable.

\begin{figure}[!t]
\begin{minipage}[t]{0.72\linewidth}
\centering
\includegraphics[width=.49\linewidth]{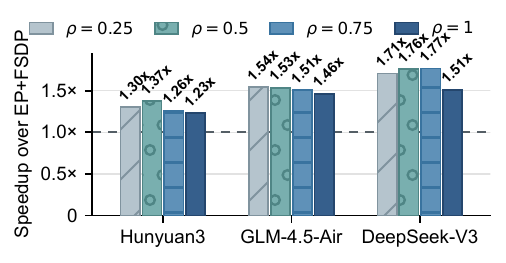}
\includegraphics[width=.49\linewidth]{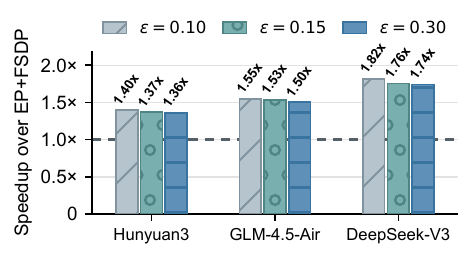}
\end{minipage}
\begin{minipage}[t]{0.27\linewidth}
\centering
\includegraphics[width=\linewidth]{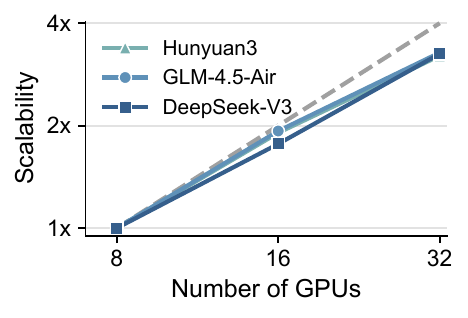}
\end{minipage}
\myvspace{-10pt}
\caption{\small{Left \& Middle: Sensitivity w.r.t. varying $\rho$ and $\varepsilon$, respectively (GLM-4.5-Air, EP=16). Right: Scalability in terms of training throughput.}}
\label{fig:eval_sensitivitiy_and_scalability}
\myvspace{-15pt}
\end{figure}

\textbf{Sensitivity.}
There are two hyper-parameters in \method{}: $\rho$ controls the maximum replicas per rank and $\varepsilon$ controls the expected workload balance ratio. 
We evaluate the sensitivity w.r.t. them individually.
Figure~\ref{fig:eval_sensitivitiy_and_scalability} provides the results with varying $\rho$ and $\varepsilon$. 
First, \method{} consistently outperforms EP+FSDP with different $\rho$. 
When $\rho=1$, the performance of \method{} drops slightly, which is due to the increased overhead of model gradient synchronization. 
However, with a moderate $\rho$ between 0.25 and 0.75, \method{}'s performance is relatively stable, demonstrating its robustness to $\rho$. 
Second, \method{} also has a robust performance w.r.t. $\varepsilon$, showcasing less than 5\% of differences in training step time when $\varepsilon$ changes. 
Thus, it does not require laborious hyper-parameter tuning.

\textbf{Scalability.}
Finally, we assess the scalability of \method{} by varying the number of GPUs to be 8, 16, and 32. 
As shown in Figure~\ref{fig:eval_sensitivitiy_and_scalability}, \method{} shows good scalability, 
improving the training throughput by 3.20-3.27$\times$ when increasing the number of GPUs from 8 to 32. 
When scaling to more GPUs, it is common to increase the data parallelism degree rather than EP, so we can apply the proposed techniques within each data parallelism group.
As a result, expanding to a larger scale would not harm \method{}'s effectiveness in reducing the cross-node communication and workload imbalance.

\section{Conclusion}
\label{sec:conclusion}

This work presents \method{}, which leverages expert co-activation to coordinate expert layout and token-expert task assignment to reduce cross-node traffic while balancing GPU workloads. Its two-stage planner uses historical co-activation and workload statistics for periodic global updates, then rebalances replicas within each node at every step. For each token, communication-aware task assignment selects the fewest remote nodes needed to serve its experts and distributes tasks among their replicas. Experimental results with three MoE models on 32 NVIDIA B200 GPUs show that \method{} achieves up to 1.53-2.41$\times$ (1.28-1.89$\times$ on average) of speedup over existing training frameworks, and substantially reduces cross-node token-transfer volume by 75.74--99.26\%.

\textbf{Limitations and possible future work.}
Due to the fact that expert selections are only available after gating, the expert layout is solved based on exponential moving average statistics rather than ad hoc statistics, and the task assignment is deduced heuristically. 
These may lead to sub-optimal solutions. Hence, a possible future direction is to refine the problem solving. 
Besides, because of the high expense of hardware resources, our evaluation is conducted on 32 GPUs. 
How \method{} behaves when it is combined with other parallelism strategies on larger clusters is worth exploring.

\bibliography{references}

@article{flux,
  title     = {{FLUX: Fast Software-based Communication Overlap On GPUs Through Kernel Fusion}},
  author    = {Chang, Li-Wen and Bao, Wenlei and Hou, Qi and Jiang, Chengquan and Zheng, Ningxin and Zhong, Yinmin and Zhang, Xuanrun and Song, Zuquan and Yao, Chengji and Jiang, Ziheng and Lin, Haibin and Jin, Xin and Liu, Xin},
  journal   = {arXiv preprint arXiv:2406.06858},
  year      = {2024},
  url       = {https://arxiv.org/abs/2406.06858}
}

@inproceedings{tamoe2022,
  title     = {{TA-MoE}: Topology-Aware Large Scale Mixture-of-Expert Training},
  author    = {Chen, Chang and Li, Min and Wu, Zhihua and Yu, Dianhai and Yang, Chao},
  booktitle = {Advances in Neural Information Processing Systems},
  volume    = {35},
  pages     = {22173--22186},
  year      = {2022},
  url       = {https://proceedings.neurips.cc/paper_files/paper/2022/hash/8b465dd58ac50e1b0b22894fd581f62f-Abstract-Conference.html}
}

@article{deepseekmoe,
  title     = {{DeepSeekMoE: Towards Ultimate Expert Specialization in Mixture-of-Experts Language Models}},
  author    = {Dai, Damai and Deng, Chengqi and Zhao, Chenggang and Xu, R. X. and Gao, Huazuo and Chen, Deli and Li, Jiashi and Zeng, Wangding and Yu, Xingkai and Wu, Y. and Xie, Zhenda and Li, Y. K. and Huang, Panpan and Luo, Fuli and Ruan, Chong and Sui, Zhifang and Liang, Wenfeng},
  journal   = {arXiv preprint arXiv:2401.06066},
  year      = {2024},
  url       = {https://arxiv.org/abs/2401.06066}
}

@article{deepseekv3,
  title     = {{DeepSeek-V3} Technical Report},
  author    = {{DeepSeek-AI}},
  journal   = {arXiv preprint arXiv:2412.19437},
  year      = {2024},
  url       = {https://arxiv.org/abs/2412.19437}
}

@article{switch,
  title     = {Switch Transformers: Scaling to Trillion Parameter Models with Simple and Efficient Sparsity},
  author    = {Fedus, William and Zoph, Barret and Shazeer, Noam},
  journal   = {Journal of Machine Learning Research},
  volume    = {23},
  number    = {120},
  pages     = {1--39},
  year      = {2022},
  url       = {https://www.jmlr.org/papers/v23/21-0998.html}
}

@inproceedings{megablocks,
  title     = {{MegaBlocks}: Efficient Sparse Training with Mixture-of-Experts},
  author    = {Gale, Trevor and Narayanan, Deepak and Young, Cliff and Zaharia, Matei},
  booktitle = {Proceedings of Machine Learning and Systems},
  volume    = {5},
  pages     = {288--304},
  year      = {2023},
  url       = {https://proceedings.mlsys.org/paper_files/paper/2023/hash/5a54f79333768effe7e8927bcccffe40-Abstract-mlsys2023.html}
}

@article{glm45,
  title     = {{GLM-4.5}: Agentic, Reasoning, and Coding ({ARC}) Foundation Models},
  author    = {{GLM-4.5 Team}},
  journal   = {arXiv preprint arXiv:2508.06471},
  year      = {2025},
  url       = {https://arxiv.org/abs/2508.06471}
}

@article{he2021fastmoe,
  title     = {{FastMoE}: A Fast Mixture-of-Expert Training System},
  author    = {He, Jiaao and Qiu, Jiezhong and Zeng, Aohan and Yang, Zhilin and Zhai, Jidong and Tang, Jie},
  journal   = {arXiv preprint arXiv:2103.13262},
  year      = {2021},
  url       = {https://arxiv.org/abs/2103.13262}
}

@inproceedings{he2022fastermoe,
  title     = {{FasterMoE}: modeling and optimizing training of large-scale dynamic pre-trained models},
  author    = {He, Jiaao and Zhai, Jidong and Antunes, Tiago and Wang, Haojie and Luo, Fuwen and Shi, Shangfeng and Li, Qin},
  booktitle = {Proceedings of the 27th ACM SIGPLAN Symposium on Principles and Practice of Parallel Programming},
  pages     = {120--134},
  year      = {2022},
  doi       = {10.1145/3503221.3508418}
}

@inproceedings{tutel,
  title     = {{Tutel}: Adaptive Mixture-of-Experts at Scale},
  author    = {Hwang, Changho and Cui, Wei and Xiong, Yifan and Yang, Ziyue and Liu, Ze and Hu, Han and Wang, Zilong and Salas, Rafael and Jose, Jithin and Ram, Prabhat and Chau, Joe and Cheng, Peng and Yang, Fan and Yang, Mao and Xiong, Yongqiang},
  booktitle = {Proceedings of Machine Learning and Systems},
  volume    = {5},
  pages     = {269--287},
  year      = {2023},
  url       = {https://proceedings.mlsys.org/paper_files/paper/2023/hash/5616d34cf8ff73942cfd5aa922842556-Abstract-mlsys2023.html}
}

@article{mixtral,
  title     = {{Mixtral of Experts}},
  author    = {Jiang, Albert Q. and Sablayrolles, Alexandre and Roux, Antoine and Mensch, Arthur and Savary, Blanche and Bamford, Chris and Chaplot, Devendra Singh and de las Casas, Diego and Bou Hanna, Emma and Bressand, Florian and Lengyel, Gianna and Bour, Guillaume and Lample, Guillaume and Renard Lavaud, L{\'e}lio and Saulnier, Lucile and Lachaux, Marie-Anne and Stock, Pierre and Subramanian, Sandeep and Yang, Sophia and Antoniak, Szymon and Le Scao, Teven and Gervet, Th{\'e}ophile and Lavril, Thibaut and Wang, Thomas and Lacroix, Timoth{\'e}e and El Sayed, William},
  journal   = {arXiv preprint arXiv:2401.04088},
  year      = {2024},
  url       = {https://arxiv.org/abs/2401.04088}
}

@inproceedings{lancet2024,
  title     = {{Lancet}: Accelerating Mixture-of-Experts Training via Whole Graph Computation-Communication Overlapping},
  author    = {Jiang, Chenyu and Tian, Ye and Jia, Zhen and Zheng, Shuai and Wu, Chuan and Wang, Yida},
  booktitle = {Proceedings of Machine Learning and Systems},
  volume    = {6},
  pages     = {74--86},
  year      = {2024},
  url       = {https://proceedings.mlsys.org/paper_files/paper/2024/hash/339caf45a6fa281cae8adc6465343464-Abstract-Conference.html}
}

@inproceedings{gshard,
  title     = {{GShard}: Scaling Giant Models with Conditional Computation and Automatic Sharding},
  author    = {Lepikhin, Dmitry and Lee, HyoukJoong and Xu, Yuanzhong and Chen, Dehao and Firat, Orhan and Huang, Yanping and Krikun, Maxim and Shazeer, Noam and Chen, Zhifeng},
  booktitle = {International Conference on Learning Representations},
  year      = {2021},
  url       = {https://arxiv.org/abs/2006.16668}
}

@inproceedings{base,
  title     = {{BASE Layers: Simplifying Training of Large, Sparse Models}},
  author    = {Lewis, Mike and Bhosale, Shruti and Dettmers, Tim and Goyal, Naman and Zettlemoyer, Luke},
  booktitle = {Proceedings of the 38th International Conference on Machine Learning},
  series    = {Proceedings of Machine Learning Research},
  volume    = {139},
  pages     = {6265--6274},
  year      = {2021},
  publisher = {PMLR},
  url       = {https://proceedings.mlr.press/v139/lewis21a.html}
}

@inproceedings{lina2023,
  title     = {Accelerating Distributed {MoE} Training and Inference with {Lina}},
  author    = {Li, Jiamin and Jiang, Yimin and Zhu, Yibo and Wang, Cong and Xu, Hong},
  booktitle = {2023 USENIX Annual Technical Conference (USENIX ATC 23)},
  pages     = {945--959},
  year      = {2023},
  url       = {https://www.usenix.org/conference/atc23/presentation/li-jiamin}
}

@inproceedings{locmoe2024,
  title     = {{LocMoE}: A Low-overhead {MoE} for Large Language Model Training},
  author    = {Li, Jing and Sun, Zhijie and He, Xuan and Zeng, Li and Lin, Yi and Li, Entong and Zheng, Binfan and Zhao, Rongqian and Chen, Xin},
  booktitle = {Proceedings of the Thirty-Third International Joint Conference on Artificial Intelligence},
  pages     = {6377--6387},
  year      = {2024},
  doi       = {10.24963/ijcai.2024/705},
  url       = {https://www.ijcai.org/proceedings/2024/705}
}

@inproceedings{li2026semantic,
  title     = {Semantic Parallelism: Redefining Efficient {MoE} Inference via Model-Data Co-Scheduling},
  author    = {Li, Yan and Zhang, Zhenyu and Wang, Zhengang and Chen, Pengfei and Zheng, Pengfei},
  booktitle = {International Conference on Learning Representations},
  year      = {2026},
  url       = {https://proceedings.iclr.cc/paper_files/paper/2026/hash/f0552f14388d95b19740dee809f5cad1-Abstract-Conference.html}
}

@inproceedings{hiermoe,
  title     = {{HierMoE}: Accelerating {MoE} Training with Hierarchical Token Deduplication and Expert Swap},
  author    = {Lin, Wenxiang and Pan, Xinglin and Zhang, Lin and Shi, Shaohuai and Wang, Xuan and Chu, Xiaowen},
  booktitle = {IEEE INFOCOM 2026 -- IEEE Conference on Computer Communications},
  pages     = {1--10},
  year      = {2026},
  doi       = {10.1109/INFOCOM59046.2026.11571570}
}

@inproceedings{liu2025netmoe,
  title     = {{NetMoE}: Accelerating {MoE} Training through Dynamic Sample Placement},
  author    = {Liu, Xinyi and Wang, Yujie and Fu, Fangcheng and Miao, Xupeng and Zhu, Shenhan and Nie, Xiaonan and Cui, Bin},
  booktitle = {International Conference on Learning Representations},
  year      = {2025},
  url       = {https://proceedings.iclr.cc/paper_files/paper/2025/hash/e0c256700465c158de71081b4cf5e8c3-Abstract-Conference.html}
}

@inproceedings{liu2026laermoe,
  title     = {{LAER-MoE}: Load-Adaptive Expert Re-layout for Efficient Mixture-of-Experts Training},
  author    = {Liu, Xinyi and Wang, Yujie and Fu, Fangcheng and Xiao, Xuefeng and Li, Huixia and Li, Jiashi and Cui, Bin},
  booktitle = {Proceedings of the 31st ACM International Conference on Architectural Support for Programming Languages and Operating Systems, Volume 2},
  pages     = {1055--1072},
  year      = {2026},
  doi       = {10.1145/3779212.3790180}
}

@inproceedings{occult2025,
  title     = {{Occult}: Optimizing Collaborative Communications across Experts for Accelerated Parallel {MoE} Training and Inference},
  author    = {Luo, Shuqing and Li, Pingzhi and Peng, Jie and Zhao, Yang and Cao, Yu and Cheng, Yu and Chen, Tianlong},
  booktitle = {Proceedings of the 42nd International Conference on Machine Learning},
  series    = {Proceedings of Machine Learning Research},
  volume    = {267},
  pages     = {41235--41253},
  year      = {2025},
  publisher = {PMLR},
  url       = {https://proceedings.mlr.press/v267/luo25f.html}
}

@inproceedings{olmoe,
  title     = {{OLMoE: Open Mixture-of-Experts Language Models}},
  author    = {Muennighoff, Niklas and Soldaini, Luca and Groeneveld, Dirk and Lo, Kyle and Morrison, Jacob and Min, Sewon and Shi, Weijia and Walsh, Pete and Tafjord, Oyvind and Lambert, Nathan and Gu, Yuling and Arora, Shane and Bhagia, Akshita and Schwenk, Dustin and Wadden, David and Wettig, Alexander and Hui, Binyuan and Dettmers, Tim and Kiela, Douwe and Farhadi, Ali and Smith, Noah A. and Koh, Pang Wei and Singh, Amanpreet and Hajishirzi, Hannah},
  booktitle = {International Conference on Learning Representations},
  year      = {2025},
  url       = {https://proceedings.iclr.cc/paper_files/paper/2025/hash/9b224ace8963c9385ad5e2b5c9039b97-Abstract-Conference.html}
}

@article{flexmoe,
  title     = {{FlexMoE}: Scaling Large-Scale Sparse Pre-Trained Model Training via Dynamic Device Placement},
  author    = {Nie, Xiaonan and Miao, Xupeng and Wang, Zilong and Yang, Zichao and Xue, Jilong and Ma, Lingxiao and Cao, Gang and Cui, Bin},
  journal   = {Proceedings of the ACM on Management of Data},
  volume    = {1},
  number    = {1},
  pages     = {1--19},
  year      = {2023},
  doi       = {10.1145/3588964}
}

@article{themis,
  title     = {{Themis}: Efficient Sparse Model Training Through Fully Sharded Sparse Data Parallelism},
  author    = {Qing, Yuhao and Zhu, Guichao and Lei, Lintian and Li, Fanxin and Zhao, Shixiong and Sun, Zekai and Guan, Xiuxian and Chen, Xusheng and Huang, Dong and Luo, Ping and Qiu, Yiming and Cui, Heming},
  journal   = {arXiv preprint arXiv:2502.02581v2},
  year      = {2025},
  url       = {https://arxiv.org/abs/2502.02581v2}
}

@inproceedings{shazeer2017outrageously,
  title     = {Outrageously Large Neural Networks: The Sparsely-Gated Mixture-of-Experts Layer},
  author    = {Shazeer, Noam and Mirhoseini, Azalia and Maziarz, Krzysztof and Davis, Andy and Le, Quoc V. and Hinton, Geoffrey E. and Dean, Jeff},
  booktitle = {International Conference on Learning Representations},
  year      = {2017},
  url       = {https://arxiv.org/abs/1701.06538}
}

@inproceedings{symi2026,
  title     = {{SYMI}: Efficient Mixture-of-Experts Training via Model and Optimizer State Decoupling},
  author    = {Skiadopoulos, Athinagoras and Zhao, Mark and Gandhi, Swapnil and Norrie, Thomas and Mukherjee, Shrijeet and Kozyrakis, Christos},
  booktitle = {23rd USENIX Symposium on Networked Systems Design and Implementation (NSDI 26)},
  pages     = {75--92},
  year      = {2026},
  url       = {https://www.usenix.org/conference/nsdi26/presentation/skiadopoulos}
}

@misc{hunyuan3,
  title        = {{Hy3}},
  author       = {{Tencent Hy Team}},
  year         = {2026},
  howpublished = {Official model release},
  url          = {https://github.com/Tencent-Hunyuan/Hy3}
}

@inproceedings{yang2026libra,
  title     = {{Libra}: Effective yet Efficient Load Balancing for Large-scale {MoE} Inference},
  author    = {Yang, Jaehoon and Kim, Yushin and Moon, Seokwon and Park, Yeonhong and Lee, Jae W.},
  booktitle = {International Conference on Learning Representations},
  year      = {2026},
  url       = {https://proceedings.iclr.cc/paper_files/paper/2026/hash/9ff1ac9a659085fed0735362cafe5e53-Abstract-Conference.html}
}

@inproceedings{smartmoe,
  title     = {{SmartMoE}: Efficiently Training Sparsely-Activated Models through Combining Offline and Online Parallelization},
  author    = {Zhai, Mingshu and He, Jiaao and Ma, Zixuan and Zong, Zan and Zhang, Runqing and Zhai, Jidong},
  booktitle = {2023 USENIX Annual Technical Conference (USENIX ATC 23)},
  pages     = {961--975},
  year      = {2023},
  url       = {https://www.usenix.org/conference/atc23/presentation/zhai}
}

@inproceedings{comet2025,
  title     = {{COMET}: Fine-grained Computation-communication Overlapping for Mixture-of-Experts},
  author    = {Zhang, Shulai and Zheng, Ningxin and Lin, Haibin and Jiang, Ziheng and Bao, Wenlei and Jiang, Chengquan and Hou, Qi and Cui, Weihao and Zheng, Size and Chang, Li-Wen and Chen, Quan and Liu, Xin},
  booktitle = {Proceedings of Machine Learning and Systems},
  volume    = {7},
  year      = {2025},
  url       = {https://proceedings.mlsys.org/paper_files/paper/2025/hash/e27ea0cd50b798ff8942caf9203f0992-Abstract-Conference.html}
}

@article{fsdp,
  title     = {{PyTorch FSDP}: Experiences on Scaling Fully Sharded Data Parallel},
  author    = {Zhao, Yanli and Gu, Andrew and Varma, Rohan and Luo, Liang and Huang, Chien-Chin and Xu, Min and Wright, Less and Shojanazeri, Hamid and Ott, Myle and Shleifer, Sam and Desmaison, Alban and Balioglu, Can and Damania, Pritam and Nguyen, Bernard and Chauhan, Geeta and Hao, Yuchen and Mathews, Ajit and Li, Shen},
  journal   = {arXiv preprint arXiv:2304.11277},
  year      = {2023},
  url       = {https://arxiv.org/abs/2304.11277}
}

@inproceedings{pit,
  title     = {{PIT: Optimization of Dynamic Sparse Deep Learning Models via Permutation Invariant Transformation}},
  author    = {Zheng, Ningxin and Jiang, Huiqiang and Zhang, Quanlu and Han, Zhenhua and Ma, Lingxiao and Yang, Yuqing and Yang, Fan and Zhang, Chengruidong and Qiu, Lili and Yang, Mao and Zhou, Lidong},
  booktitle = {Proceedings of the 29th Symposium on Operating Systems Principles},
  pages     = {331--347},
  year      = {2023},
  doi       = {10.1145/3600006.3613139},
  url       = {https://doi.org/10.1145/3600006.3613139}
}

@inproceedings{expertchoice,
  title     = {{Mixture-of-Experts with Expert Choice Routing}},
  author    = {Zhou, Yanqi and Lei, Tao and Liu, Hanxiao and Du, Nan and Huang, Yanping and Zhao, Vincent and Dai, Andrew M. and Chen, Zhifeng and Le, Quoc V. and Laudon, James},
  booktitle = {Advances in Neural Information Processing Systems},
  volume    = {35},
  pages     = {7103--7114},
  year      = {2022},
  url       = {https://proceedings.neurips.cc/paper_files/paper/2022/hash/2f00ecd787b432c1d36f3de9800728eb-Abstract-Conference.html}
}

@manual{nvidia_dgxb200,
  author = {{NVIDIA}},
  title = {{NVIDIA DGX B200 User Guide: Introduction to NVIDIA DGX B200 Systems}},
  year = {2026},
  note = {Accessed September 16, 2026},
  url = {https://docs.nvidia.com/dgx/dgxb200-user-guide/introduction-to-dgxb200.html}
}

@misc{kimik3,
  title        = {{Kimi K3}: Open Frontier Intelligence},
  author       = {{Moonshot AI}},
  year         = {2026},
  howpublished = {Official model release},
  url          = {https://github.com/MoonshotAI/Kimi-K3}
}

@article{deepseekv4,
  title   = {{DeepSeek-V4}: Towards Highly Efficient Million-Token Context Intelligence},
  author  = {{DeepSeek-AI}},
  journal = {arXiv preprint arXiv:2606.19348},
  year    = {2026},
  url     = {https://arxiv.org/abs/2606.19348}
}

@article{glm5,
  author  = {{GLM-5 Team}},
  title   = {{GLM-5}: From Vibe Coding to Agentic Engineering},
  journal = {arXiv preprint arXiv:2602.15763},
  year    = {2026},
  url     = {https://arxiv.org/abs/2602.15763}
}

@misc{deepep2025,
  title        = {{DeepEP}: An Efficient Expert-Parallel Communication Library},
  author       = {Zhao, Chenggang and Zhou, Shangyan and Zhang, Liyue and Deng, Chengqi and Xu, Zhean and Liu, Yuxuan and Yu, Kuai and Li, Jiashi and Zhao, Liang},
  year         = {2025},
  howpublished = {GitHub},
  url          = {https://github.com/deepseek-ai/DeepEP}
}

@article{scip,
  title   = {{SCIP}: Solving Constraint Integer Programs},
  author  = {Achterberg, Tobias},
  journal = {Mathematical Programming Computation},
  volume  = {1},
  number  = {1},
  pages   = {1--41},
  year    = {2009},
  doi     = {10.1007/s12532-008-0001-1}
}

@article{raffel2020t5,
  title   = {Exploring the Limits of Transfer Learning with a Unified Text-to-Text Transformer},
  author  = {Raffel, Colin and Shazeer, Noam and Roberts, Adam and Lee, Katherine and Narang, Sharan and Matena, Michael and Zhou, Yanqi and Li, Wei and Liu, Peter J.},
  journal = {Journal of Machine Learning Research},
  volume  = {21},
  number  = {140},
  pages   = {1--67},
  year    = {2020},
  url     = {https://www.jmlr.org/papers/v21/20-074.html}
}
\bibliographystyle{iclr2027_conference}

\clearpage
\appendix
\section{More Experimental Details}
\label{sec:appendix-eval}

\paragraph{Hardware environments.}
All experiments are conducted on 4 GPU servers, with each server containing 8 NVIDIA B200 GPUs, totaling 32 GPUs. 
Each GPU is equipped with 192 GB of HBM3e memory that provides up to $8$ TB/s memory access bandwidth. 
GPUs within each node are interconnected by NVLink with $1.8$ TB/s of bandwidth. 
For cross-node communication, each node is equipped with 8 ConnectX-7 InfiniBand NICs, providing up to $400$ GB/s of aggregate bandwidth.

\paragraph{Additional workload configurations.}
Table~\ref{tab:experiment-configuration}
lists more details about the evaluated models, 
and the default training configurations used in our experiments. 

\begin{table}[H]
  \centering
  \caption{\small{Detailed model and training configurations.}}
  \label{tab:experiment-configuration}
  \small
  \setlength{\tabcolsep}{6pt}
  \begin{tabular*}{\linewidth}{@{\extracolsep{\fill}}lccccccc@{}}
    \toprule
    Model & \specialcell{Hidden\\Dim} & \specialcell{FFN\\Dim} & \specialcell{Num.\\Experts} & \specialcell{Selected\\Experts} & \specialcell{Attn.\\heads} & \specialcell{Num.\\Layers} & \specialcell{Total Params \&\\Activated Params per Token} \\
    \midrule
    Hunyuan3 & 4096 & 1536 & 192 & 8 & 64 & 16 & 60B-A5B \\
    GLM-4.5-Air & 4096 & 1408 & 128 & 8 & 96 & 45 & 106B-A12B \\
    DeepSeek-V3 & 7168 & 2048 & 256 & 8 & 128 & 12 & 140B-A8B \\
    \bottomrule
  \end{tabular*}
  \par\smallskip
  \begin{tabular*}{\linewidth}{@{\extracolsep{\fill}}lrr@{}}
    \toprule
    Configuration & EP16 & EP32 \\
    \midrule
    Sequence length & 8192 & 8192 \\
    Micro-batch size & 2 & 2 \\
    Local batch & 4 & 4 \\
    Global batch & 64 & 128 \\
    \bottomrule
  \end{tabular*}
\end{table}

\section{MILP Reformulation}
\label{sec:appendix-milp}

In Section~\ref{sec:mone-layout}, we mention that 
Equation~\eqref{eq:mone-planning} can be re-written as 
a mixed-integer linear programming (MILP) problem. 
Below we explain how it is reformulated.

Equation~\eqref{eq:mone-planning} chooses both the expert layout and
how to divide each expert's estimated work among its replicas. 
We express these decisions by introducing 
auxiliary variables: $x_{er}\in\{0,1\}$ indicates whether rank $r$ holds a replica of expert $e$, 
and $l_{er}\ge0$ represents the work assigned to that replica. 
Thus, $\mathcal P_r=\{e:x_{er}=1\}$ and
$l_{er}$ represents $w^\prime_{er}$ in
Equation~\eqref{eq:mone-planning}. 
Let $\mathcal R_n=\{r:\texttt{N}(r)=n\}$ be the ranks on node $n$.
The variable $u_{en}\in[0,1]$ indicates whether node $n$ holds at
least one replica of expert $e$. For each expert pair $i<j$ and node
$n$, the variable $z_{ijn}\in[0,1]$ indicates whether both experts
have replicas on that node.
The resulting mixed-integer linear programing problem is
\begin{equation}
\begin{aligned}
  \argmax_{\{x_{er}\},\,\{u_{en}\},\,\{z_{ijn}\},\,\{l_{er}\}}\quad
    & \textstyle\sum_n\sum_{i<j}\hat c_{ij}z_{ijn}\\
  \text{s.t.}\quad
    & x_{er}\le u_{en},
      \, \forall e,\ n,\ r\in\mathcal R_n,
    \quad && u_{en}\le\textstyle\sum_{r\in\mathcal R_n}x_{er},
      \, \forall e,\ n,\\
    & z_{ijn}\le u_{in},\quad z_{ijn}\le u_{jn},
      \, \forall i<j,\ n,
    && z_{ijn}\ge u_{in}+u_{jn}-1,
      \, \forall i<j,\ n,\\
    & x_{er}\in\{0,1\},
      \, \forall e,\ r,
    && u_{en}\in[0,1],
      \, \forall e,\ n,\\
    & z_{ijn}\in[0,1],
      \, \forall i<j,\ n,
    && \textstyle\sum_r x_{er}\ge1,
      \, \forall e, \\
    & \textstyle\sum_e x_{er}\le(1+\rho)E/G,
      \, \forall r,
    && \textstyle\sum_r l_{er}=\hat w_e,
      \, \forall e, \\
    & 0\le l_{er}\le\hat w_e x_{er},
      \, \forall e,\ r,
    && \textstyle\sum_e l_{er}\le(1+\varepsilon)KT/G,
      \, \forall r,
\end{aligned}
\label{eq:appendix-milp}
\end{equation}
The first two constraints force $u_{en}=1$ exactly when node $n$
holds a replica of expert $e$. The next two constraints then force
$z_{ijn}=1$ exactly when node $n$ holds replicas of both experts.
Because $x_{er}$ is binary, both $u_{en}$ and $z_{ijn}$ need only
continuous bounds. Therefore, it counts $\hat c_{ij}$
once for each node holding both experts, even if their replicas
reside on different ranks of that node, exactly as in
Equation~\eqref{eq:mone-planning}.
The remaining constraints enforce expert coverage and the per-rank
replica budget, conserve each expert's estimated work, and prevent a
rank without its replica from receiving that work. The last workload
inequality caps each rank's planned work at the same threshold as
Equation~\eqref{eq:mone-planning}. Hence, every feasible solution maps
to a feasible placement and workload division in the original
planning problem, and vice versa, with the same objective value. Only
the $EG$ placement variables $x_{er}$ need to be integer.
Consequently, Equation~\eqref{eq:appendix-milp} is an MILP reformulation of Equation~\eqref{eq:mone-planning}.

\end{document}